# Slamming of a Breaking Wave on a Wall

*Jian-Jun Shu*
School of Mechanical & Aerospace Engineering, Nanyang Technological University, 50 Nanyang Avenue, Singapore 639798

**ABSTRACT**

This paper is intended to study impact forces of breaking waves on a rigid wall based on a nonlinear potential-flow theory. This is a model problem for some technologically important design issues such as the impact of breaking waves on ships, coastal and offshore structures. We are interested in the short-time successive triggering of nonlinear effects using a small-time expansion. The analytical solutions for the impact force on a rigid wall and the free-surface profile are derived.

**KEY WORDS:** Breaking wave, plunging wave, impulsive pressure

## INTRODUCTION

There is a long history of experimental and theoretical studies to determine impact forces acting on a rigid wall, which is suddenly started from rest and made to move towards a fluid taper. The problem is motivated by the impact of breaking waves on ships, coastal and offshore structures, which is one of the most severe environmental loads on structures. The impact due to a breaking wave striking a wall is of high intensity and short duration. This is attributed to the direct collision between a fluid and a wall surface. The direct collision of a breaking wave with a wall generates an impulsive pressure on the wall. This is similar to the problem of initial-stage water impact. Unfortunately, existing wave theories based on small- and finite-amplitude assumptions cannot be directly adopted to evaluate the breaking wave force on a wall due to the highly nonlinear and transient nature of the problem.

In reviewing the previous studies, one of the most important and unresolved questions is how the initial stage of the breaking wave impingement on the wall can be properly characterized and simulated. Cumberbatch (1960) considered the case of symmetric normal impact of a water wedge on a wall and Zhang et al. (1996) extended his work to an oblique impact. These two works stemmed from an *ad hoc* assumption on the free-surface profiles close to the wall: in Cumberbatch (1960), a linear function was assumed, while in Zhang et al. (1996), an exponential function was used.

In the present study the free-surface profiles are analytically determined without prescribed functions. Effects of gravity, viscosity and surface tension can be neglected since inertia forces are dominant during the small-time impact process. The essential mechanism involved in the impact process can be described by the theoretical treatment of potential flow. A small portion of the breaker tip is initially cut off to produce a finite wetted area on the wall and a high spike in the consequent impact results from an acceleration of water towards the wall. We are interested here in the short-time successive triggering of nonlinear effects using a small-time expansion of the full, nonlinear initial/boundary value problem. The leading small-time expansion is taken to include the accelerating effect. The analytical solutions for the hydrodynamic force on a wall and the free-surface profile are derived. It is worth to mention to this end that the technique proposed by Chwang (1978) and extended by Chwang (1983), Liu (1986) and King & Needham (1994) to investigate the earthquake effect on dams has been adopted here in the mathematical treatment to the present problem although physical settings are different.

## GOVERNING EQUATIONS

We consider a rigid horizontal wall, being suddenly started from rest and made to move vertically with constant acceleration $a_0$ towards a two-dimensional fluid taper with semi-angle $\alpha\pi\,(0<\alpha<1/2)$. A definition sketch of the flow is shown in Figure 1. The axis of the fluid taper is perpendicular to the wall. Let us nondimensionalize time $t$ by $(L/a_0)^{1/2}$, distance $(x,y)$ by $L$, velocity $(u,v)$ by $(a_0 L)^{1/2}$, pressure $p$ by $\rho a_0 L$, where $L$ is the wetted wall semi-length when the breaking wave just touches the wall at time $t=0$ and $\rho$ is the density of the fluid. A mathematical statement of the above problem can now be written as

$$\frac{\partial u}{\partial x}+\frac{\partial v}{\partial y}=0, \qquad (1)$$

$$\frac{\partial u}{\partial t}+u\frac{\partial u}{\partial x}+v\frac{\partial u}{\partial y}=-\frac{\partial p}{\partial x}, \qquad (2)$$

$$\frac{\partial v}{\partial t}+u\frac{\partial v}{\partial x}+v\frac{\partial v}{\partial y}=-\frac{\partial p}{\partial y}. \qquad (3)$$

For negative time $t<0$ everything is at rest,

$$u=v=0, \eta=0 \text{ for } t<0, \qquad (4)$$

where $\eta$ is the free surface ``elevation'' in the $x$ direction beyond the undisturbed surface. On the surface, the kinematic and dynamic boundary conditions require

$$u = \frac{\partial \eta}{\partial t} + v\frac{\partial \eta}{\partial y},\ p=0 \text{ on } x=1+y\tan(\alpha\pi)+\eta(y,t). \qquad (5)$$

On the wall surface, the normal velocity of fluid particles must be the same as that of the wall at all time

$$v = a_0 t \text{ on } y = a_0 t^2/2. \qquad (6)$$

On the axis of the fluid taper, the normal velocity of the fluid must vanish from consideration of symmetry about the axis of the fluid taper,

$$u=0 \text{ on } x=0. \qquad (7)$$

The pressure vanishes at infinity,

$$p \to 0 \text{ as } y \to \infty. \qquad (8)$$

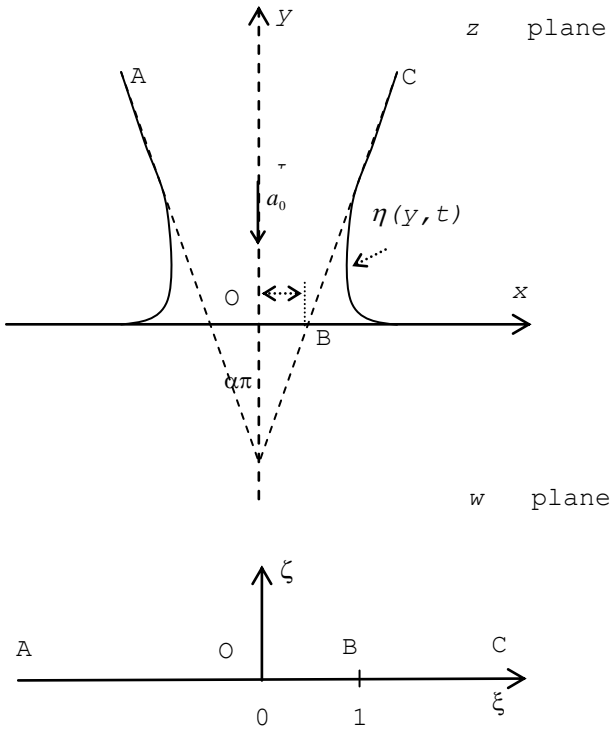

Figure 1. Physical $z$ plane is conformally mapped onto the upper half of the $w$ plane.

The solution domain for this set of equations (1)-(3) with conditions (4)-(8) is unknown at this stage of the analysis but is conveniently described as

$$\{(x,y,t): 0 \le x \le 1+y\tan(\alpha\pi)+\eta(y,t),\ a_0 t^2/2 \le y < \infty,\ 0 \le t < \infty\}.$$

## MATHEMATICAL ANALYSIS

The full nonlinear initial/boundary value problem consists of equations (1)-(3) with conditions (4)-(8). These equations are solved analytically by employing a small-time expansion. We assume that

$$u(x,y,t) = u_1(x,y)t + O(t^2),\ v(x,y,t) = v_1(x,y)t + O(t^2), \qquad (9)$$

$$p(x,y,t) = p_0(x,y) + O(t),\ \eta(y,t) = \eta_2(y)t^2 + O(t^3). \qquad (10)$$

The leading-order equations are

$$\frac{\partial u_1}{\partial x} + \frac{\partial v_1}{\partial y} = 0,\ u_1 = -\frac{\partial p_0}{\partial x},\ v_1 = -\frac{\partial p_0}{\partial y} \qquad (11)$$

subject to the conditions

$$u_1 = 2\eta_2,\ p_0 = 0 \text{ on } x = 1+y\tan(\alpha\pi), \qquad (12)$$

$$v_1 = a_0 \text{ on } y=0, \qquad (13)$$

$$u_1 = 0 \text{ on } x=0, \qquad (14)$$

$$p_0 \to 0 \text{ as } y \to \infty. \qquad (15)$$

It is clear that pressure $p_0$ satisfies the Laplace equation

$$\frac{\partial^2 p_0}{\partial x^2} + \frac{\partial^2 p_0}{\partial y^2} = 0. \qquad (16)$$

Introducing a complex-conjugate function $q_0$ with respect to $p_0$, we can construct an analytic function

$$f_0(z) \equiv p_0 + iq_0,\ z = x+iy. \qquad (17)$$

As shown in Figure 1, the conformal mapping

$$z = 1 + \frac{\alpha\Gamma(\alpha)e^{i(1/2-\alpha)\pi}}{\sqrt{\pi}\Gamma(1/2+\alpha)} \int_1^w \tau^{-1/2}(\tau-1)^{\alpha-1/2} d\tau \qquad (18)$$

given by the Schwarz-Christoffel transformation, maps the upper half of the $w$ plane ($w = \xi + i\zeta$) onto the region occupied by the fluid. Here $\Gamma$ is the Gamma function defined by

$$\Gamma(w) = \int_0^\infty \tau^{w-1} e^{-\tau} d\tau.$$

Function $f_0$ is also analytic in the transformed variable $w$. On the free surface, which corresponds to $\xi > 1$ on the positive real axis, $p_0$ vanishes. On the axis of the fluid taper, which corresponds to the negative real axis in the $w$ plane, we have $\partial p_0/\partial n = 0$, which means that $q_0$ is a constant. Without loss of generality, we may assume $q_0 = 0$ for $\xi < 0$ along the negative real axis. On the wall surface, which corresponds to the line segment $0 < \xi < 1$, we take $\partial p_0/\partial n = a_0$. Therefore, along the real axis in the $w$ plane, we have

$$\text{Im}(f_0) = 0 \text{ on } -\infty < \xi < 0, \quad (19)$$

$$\text{Re}\left(\frac{\partial f_0}{\partial n}\right) = a_0 \text{ on } 0 < \xi < 1, \quad (20)$$

$$\text{Re}(f_0) = 0 \text{ on } 1 < \xi < \infty. \quad (21)$$

If $s(\xi)$ measures the distance from point **B** in Figure 1 to any point on the wall surface, the Cauchy-Riemann equations give

$$\text{Im}(f_0) = 0 \text{ on } -\infty < \xi < 0, \quad (22)$$

$$\text{Im}(f_0) = a_0 s(\xi) \text{ on } 0 < \xi < 1, \quad (23)$$

$$\text{Re}(f_0) = 0 \text{ on } 1 < \xi < \infty, \quad (24)$$

where the distance $s(\xi)$ is given by (18) as

$$s(\xi) = \frac{\alpha \Gamma(\alpha)}{\sqrt{\pi}\Gamma(1/2+\alpha)} \int_\xi^1 \tau^{-1/2}(1-\tau)^{\alpha-1/2} d\tau \text{ on } 0 < \xi < 1. \quad (25)$$

If we introduce a new analytic function $g_0(w)$ by

$$g_0(w) = (1-w)^{-1/2} f_0(w), \quad (26)$$

the boundary conditions for $g_0(w)$ are unmixed

$$\text{Im}(g_0) = 0 \text{ on } -\infty < \xi < 0, \quad (27)$$

$$\text{Im}(g_0) = a_0(1-\xi)^{-1/2} s(\xi) \text{ on } 0 < \xi < 1, \quad (28)$$

$$\text{Im}(g_0) = 0 \text{ on } 1 < \xi < \infty. \quad (29)$$

The analytic function $g_0(w)$ which is regular in the upper half $w$ plane and vanishes at infinity can be obtained from the Schwarz integral formula

$$g_0(w) = \frac{1}{\pi} \int_{-\infty}^\infty \frac{\text{Im}(g_0)}{\xi - w} d\xi. \quad (30)$$

Substituting (26) -- (29) into (30), we have

$$f_0(w) = \frac{a_0(1-w)^{1/2}}{\pi} \int_0^1 \frac{s(\xi)}{(1-\xi)^{1/2}(\xi-w)} d\xi. \quad (31)$$

The impact pressure on the wall is the real part of $f_0(w)$ for $0 < \xi < 1$. Using (25) and integrating by parts, we have on $0 < \xi < 1$

$$P_0(\xi) = \text{Re}(f_0|_{\zeta=0}) = \frac{a_0\Gamma(1+\alpha)}{\pi^{3/2}\Gamma(1/2+\alpha)} \\ \times \fint_0^1 \tau^{-1/2}(1-\tau)^{\alpha-1/2} \ln\left|\frac{(1-\tau)^{1/2} + (1-\xi)^{1/2}}{(1-\tau)^{1/2} - (1-\xi)^{1/2}}\right| d\tau, \quad (32)$$

where $\fint$ denotes the Cauchy principal value. It is better to express the above formula in the form that is suitable for easy computation. Differentiating (32) with respect to $\xi$, we obtain

$$\frac{dP_0}{d\xi} = \frac{a_0\Gamma(1+\alpha)}{\pi^{3/2}(1-\xi)^{1/2}\Gamma(1/2+\alpha)} \fint_0^1 \frac{\tau^{-1/2}(1-\tau)^\alpha}{\tau - \xi} d\tau \\ \text{on } 0 < \xi < 1. \quad (33)$$

The integral on the right-hand side of (33) can be obtained by contour integration. Thus

$$\frac{dP_0}{d\xi} = \frac{a_0\Gamma(1+\alpha)\cot(\alpha\pi)}{\pi^{3/2}(1-\xi)^{1/2}\Gamma(1/2+\alpha)}\left[\frac{\pi(1-\xi)^\alpha}{\xi^{1/2}} - \int_0^\infty \frac{(1+\tau)^\alpha}{\tau^{1/2}(\tau+\xi)} d\tau\right] \\ \text{on } 0 < \xi < 1. \quad (34)$$

Integrating (34), we have

$$P_0(\xi) = \frac{8a_0\Gamma(1+\alpha)\cot(\alpha\pi)}{\pi^{3/2}(1-\xi)^{1/2}\Gamma(1/2+\alpha)} \\ \times \int_0^{\pi/2} \frac{\xi^{1/2}}{(\tan^2\theta + \xi)^{1/2-\alpha}} \frac{\theta d\theta}{\sin(2\theta)(\tan\theta)^{2\alpha}} - a_0 s(\xi) \text{ on } 0 < \xi < 1. \quad (35)$$

From boundary conditions (12) to (15), we have

$$\left.\frac{\text{Im}(\partial f_0/\partial w)}{\text{Im}(\partial z/\partial w)}\right|_{\zeta=0} = 2\eta_2(\xi) \text{ on } \xi > 1. \quad (36)$$

After some mathematical manipulation, we obtain

$$\eta_2(\xi) = \frac{a_0\pi^{1/2}(1/2+\alpha)\Gamma(1+\alpha)B(1/2+\alpha;-\alpha;1/\xi)}{2(\xi-1)\sin[(1/2-\alpha)\pi]\Gamma(3/2+\alpha)} \text{ on } \xi > 1. \quad (37)$$

where $B$ is the incomplete Beta function

$$B(\nu;\mu;\omega) = \int_0^\omega \tau^{\nu-1}(1-\tau)^{\mu-1} d\tau.$$

Impact free surface profiles for different semi-angle tapers are shown in Figures 2. It has been found that the free-surface profile $\eta_2(y)$ close to the wall is proportional to $y^{2/(1-2\alpha)}$, which is neither linear in Cumberbatch's assumption (1960) nor exponential in Zhang et al.'s assumption (1996). Figure 3 shows that the impact hydrodynamic pressure increases as $\xi$ increases, whereas the pressure decreases as $\alpha$ increases. Therefore, the pressure distribution for $\alpha \to 0$ is the maximum envelope of all pressure distributions. The maximum pressure always occurs on the free surface. It is clear that negative impact pressure appears near the axis of the fluid taper and consequent cavitation will be generated in that region.

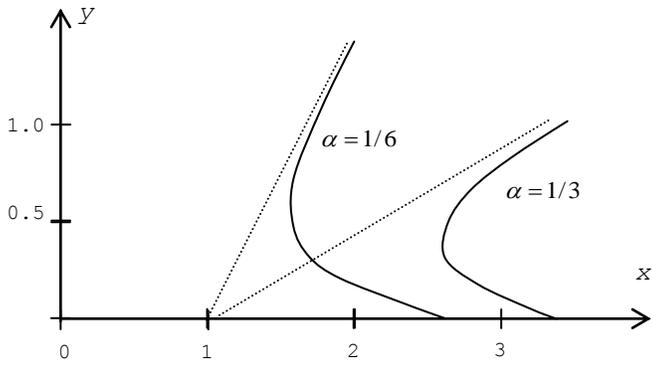

Figure 2. Impact free surface shapes $\eta_2(\xi)/a_0$ for various semi-angle $\alpha$ tapers.

$\alpha = 1/6$

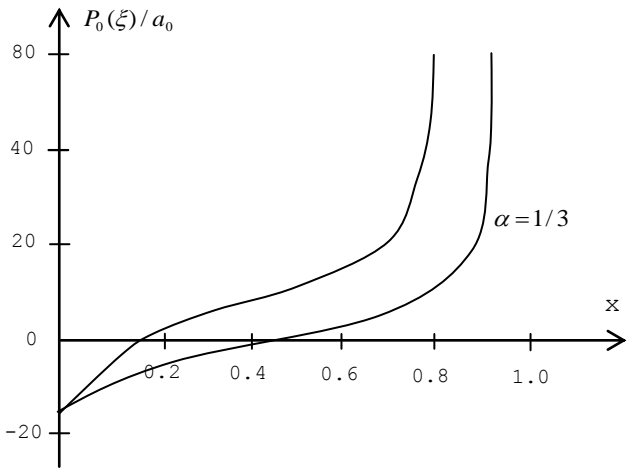

Figure 3. Impact pressure distributions $P_0(\xi)/a_0$ on the wall for various semi-angle $\alpha$ tapers.

## CONCLUSIONS

An analytical approach is pursued to study the impact force of a breaking wave on a rigid wall. The initial stage of the impact is characterized by an impact of a two-dimensional liquid taper acting on the wall with a prescribed acceleration. The problems of the impact forces of breaking waves impingement on the wall and the free-surface profile have been solved analytically by using a small-time expansion. Explicit analytical formulae for evaluating the impact pressure and the free-surface profile have been given. It has been found that the free-surface profile close to the wall is neither linear in Cumberbatch's assumption (1960) nor exponential in Zhang et al.'s assumption (1996).